\documentclass[twocolumn,prl,showpacs,footinbib]{revtex4}
\usepackage{graphicx}
\usepackage{amssymb}
\usepackage{epsfig}

\begin{document}

\title{Curvature condensation and bifurcation in an elastic shell}
\author{Moumita Das$^1$}
\author{Ashkan Vaziri$^1$}
\author{Arshad Kudrolli$^2$}
\author{L. Mahadevan$^1$}
\email{lm@deas.harvard.edu}
\affiliation{$^1$ Division of Engineering and Applied Sciences, Harvard University, Cambridge, MA 02138\\
$^2$ Department of Physics, Clark University, Worcester, MA 01610}
%\draft

\begin{abstract}
We study the formation and evolution of  localized geometrical defects in an indented cylindrical elastic shell using a combination of experiment and numerical simulation. We find that as a symmetric localized indentation  on a semi-cylindrical shell increases, there is a transition from a global mode of deformation to a localized one which leads to the condensation of curvature along a symmetric parabolic crease. This process introduces a soft mode in the system, converting a load-bearing structure into a hinged, kinematic mechanism. Further indentation leads to  twinning wherein the parabolic crease bifurcates into two creases that move apart on either side of the line of symmetry. A qualitative theory captures the main features of the phenomena and leads to sharper questions about the nucleation of these defects.

\end{abstract}
\date{\today}
\maketitle 

%\section{Introduction}

\begin{figure}
\includegraphics[width=5cm,height=4cm]{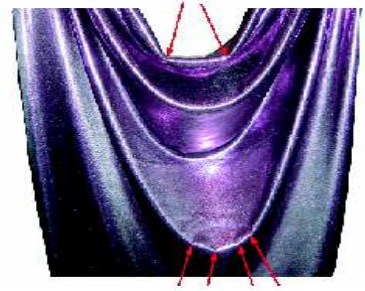}
\caption{\label{drape} Condensation of curvature in a heavy drape. The smooth catenary-like folds with a single direction of localized curvature coexist with regions of curvature condensation, indicated by arrows, where curvature is strongly focused.}
\end{figure}

The formation of defects in continuum physics  has been the subject of long-standing investigation in many ordered and partially ordered bulk condensed matter systems such as crystals, liquid crystals and various quantum ``super''-phases \cite{Nelson:2002}. In low-dimensional systems, the formation and evolution of defects involves an extra level of complexity due to the interplay between geometry and physics and remains an active area of research. Even in simple  physical systems such as the mundane paper we write on and the textile sheets we wear daily, the visual effects of these defects is arresting; in Fig.\ref{drape}  we see  a  heavy hanging drape  with curved catenary-like wrinkles that themselves have  crease-like anisotropic defects (see arrows) on scales much smaller than the crease but much larger than the thickness of the drape. This hierarchy in structural scales is common in many elastic systems  and naturally suggests the question of how such structures form. While the actual shape, response and stability of these structures has been the subject of much study recently \cite{Lobkovsky:97}, \cite{Cerda:98}, \cite{Boudaoud:00}, \cite{Cerda:05}, following the early work of Pogorelov \cite{Pogorelov}, the process by which they arise from a featureless sheet is essentially unknown.

\begin{figure}
\includegraphics[width=8cm,height=12cm]{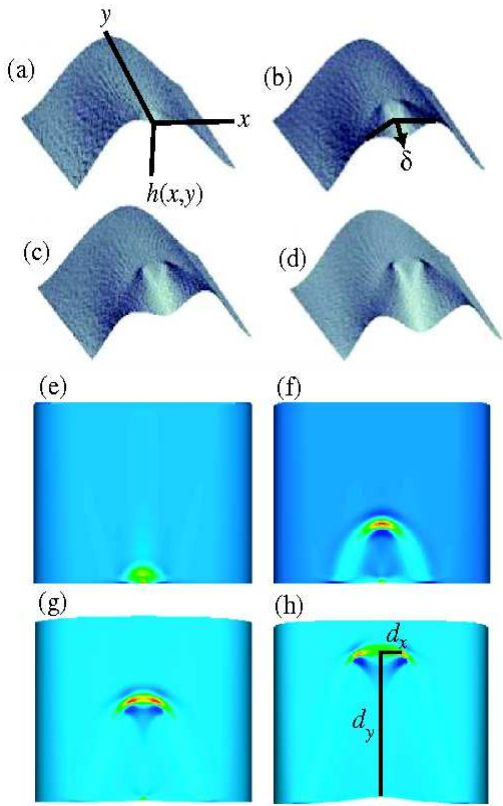}
\caption{\label{curvevolve} (Color online) 
Curvature condensation in an indented semi-cylinder. 
Figs (a) to (d) correspond to the experimentally determined deformation of the sheet with the colormap representing curvature $h_{,yy}$ along the axis of cylindrical symmetry.  
Figs (e) to (h) show the evolution of the  curvature  $h_{,yy}$  with increasing indentation $\delta$ (sheet viewed from the top) , for $t/R=0.001$ determined using numerical simulations (see text); red corresponds to high curvatures, blue to low.. The Gauss curvature $\kappa_G \sim \frac{1}{R}  h_{yy}$ shows the same qualitative trends in both cases.
}
\end{figure}

To address this question in the simple context of an everyday example, a thin mylar sheet is bent into a half-cylindrical elastic shell (thickness $t$, radius $R$ and length $L$, $t \ll R<L$; $R/t \sim 100$) that is clamped along its lateral edges and indented at one end as shown in Fig. \ref{curvevolve} (a-d), a system on which preliminary investigations were  conducted a while ago \cite{Calladine:94}.  A point probe that is free to slip on the mylar is then used to indent the sheet at one end along the axis of symmetry.  The shape of the sheet is then  reconstructed using laser aided tomography \cite{Blair:2004}, wherein a laser sheet is used to interrogate the surface and determine its height.  A CCD camera with a resolution of 1024 $\times$ 768 pixels is used to image the reflected light and the laser sheet is rotated with a stepper motor which allows the entire surface to be scanned.   

For small values of the indentation $\delta$, the shell deforms strongly in the immediate neighborhood of the edge, and the deformation decays monotonically on a scale comparable to the cylinder radius (Fig.\ref{curvevolve}a). As the indentation is increased further, we find that  a localized zone of high curvature  appears along the axis of symmetry, but at  a distance $d_y$ away from the point of indentation (Fig.\ref{curvevolve}b). This ``condensation'' of curvature along a localized parabolic defect in an elastic sheet is similar to that at the focus of a developable cone \cite{Cerda:98,Cerda:05}. For larger values of the indentation, if $R/t$ is sufficiently large, the parabolic crease of localized curvature bifurcates into two such creases that move away symmetrically from the axis of the cylinder, in a manner analogous to defect twinning (Fig.\ref{curvevolve}c).  Finally, for even larger values of the indentation, the defects stop moving along the axis of the cylinder, and a large region of the shell folds inwards (Fig.\ref{curvevolve}d).  For long cylinders, the only intrinsic dimensionless parameter is the ratio $R/t$; there is effectively no dependence on material parameters since the long wavelength modes of deformation of a thin shell associated with bending and stretching are both characterized by the Young's modulus $E$  and thickness $t$ \footnote{There is a dependence on the Poisson ratio, but this is weak and we will ignore it here.}. This makes the problem particularly attractive since geometry is at the heart of all the observed phenomena. 

Understanding the experimental results requires the solution of the nonlinear equations for the deformations of shells \cite{Antman}. However, these are analytically insoluble except in some special cases involving axisymmetric geometries and deformations. Therefore, we resort to a numerical investigation, and complement this with simple scaling analyses to tease out the qualitative aspects of our results.  A commercially available finite element code ABAQUS is used to minimizes the elastic energy of the shell (made of a material that is assumed to be linearly elastic, with  Young's modulus $E =100$ MPa and Poisson's ratio $\nu = 0.3$) with an energy density
\begin{eqnarray} 
 U &=& \frac{Et}{2(1-\nu^2)} \left[({\epsilon_1}+ {\epsilon_2})^2 - 2 (1-\nu )(\epsilon_1 \epsilon_2 - \gamma^2) \right] \nonumber\\ 
&+&\frac{Et^3}{24(1-\nu^2)} \left[({\kappa_1}+ {\kappa_2})^2 - 2 (1-\nu)( \kappa_1 \kappa_2 - \tau^2) \right] \nonumber. 
 \end{eqnarray}
Here the first line characterizes the energy associated with in-plane deformations (with strains $\epsilon_1, \epsilon_2, \gamma$ ) and the second line accounts for the energy associated with out-of-plane deformations ($\kappa_1, \kappa_2, \tau$ are the curvatures and twist relative to the undeformed tube).  Four-node, quadrilateral shell elements with reduced integration and a large-deformation formulation  were used in the calculations. 

\begin{figure}
\includegraphics[width=7cm]{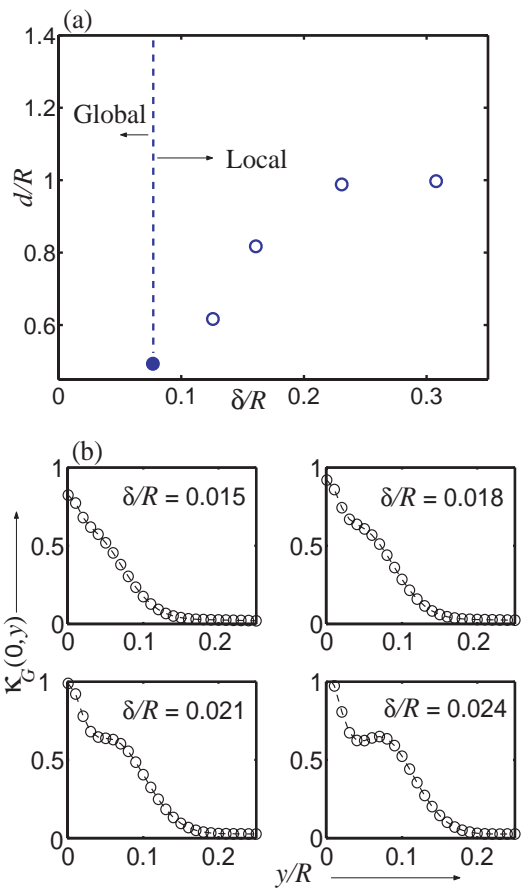}
\caption{\label{globaltolocal} Global and local modes of deformation. (a) Experimentally determined  location of the condensate as a function of indentation depth, for a  mylar sheet ($t/R\sim .01$).  Filled symbols represent the  condensate (before bifurcation) and open symbols represent the twinned state (after bifurcation). (b) We can understand this transition by probing the Gauss curvature along the axis of symmetry $\kappa_{G} (0,y) \sim \frac{1}{R}h_{,yy}(0,y)$
shown in determined numerically. We see the transition from a global mode with just a single curvature maximum (at the point of indentation) to a localized mode of deformation with a secondary maximum in the curvature that arises when the indentation $\delta$ is larger than some threshold that is a function solely of the cylinder geometry. Here $t/R=0.001$.}
\end{figure}

The results are shown in (Fig. \ref{curvevolve} e-h): as the indentation is increased, the curvature condenses at a distance away from the edge, and this ``condensate'' bifurcates into two upon further indentation. To understand these transitions qualitatively, we note that a  complete cylindrical shell  that is pinched gently at one end responds by deforming globally in a nearly inextensional mode. Although a semi-cylinder clamped along its lateral edges also does the same  when indented weakly,  this global mode is screened by the lateral clamps. Strong indentation then leads to a transition from the global mode to a localized mode wherein the curvature localizes in a very small region at a distance $d_y$ from the edge.  In Fig.\ref{globaltolocal}a, we show that this transition  arises  only when the indentation is larger than a threshold. To further interrogate this phenomenon, in  Fig.\ref{globaltolocal}b, we plot the Gauss curvature along the axis of symmetry of the cylinder as a function of the indentation. For small indentations, we find that the Gauss curvature along the axis of symmetry decreases monotonically from the zone of indentation as shown in Fig.\ref{globaltolocal}b. However as the indentation is increased, the Gauss curvature develops an inflection point and eventually a secondary maximum, where the condensate forms. This condensation is crucially dependent on the two-dimensionality of the problem, and is not seen in one dimensional elastic systems \cite{Thompson:99}. There are rough analogies to a phase transition when interpreted in terms of the delocalized Gauss curvature and its localized condensate. Indeed the infinite-dimensional cylindrical structure with finite stiffness becomes a finite-dimensional mechanism, acquiring a new soft degree of freedom with a very low stiffness due to the formation of a hinge at the location of high curvature, which is akin to a condensed phase.

\begin{figure}
\includegraphics[width=7.5cm]{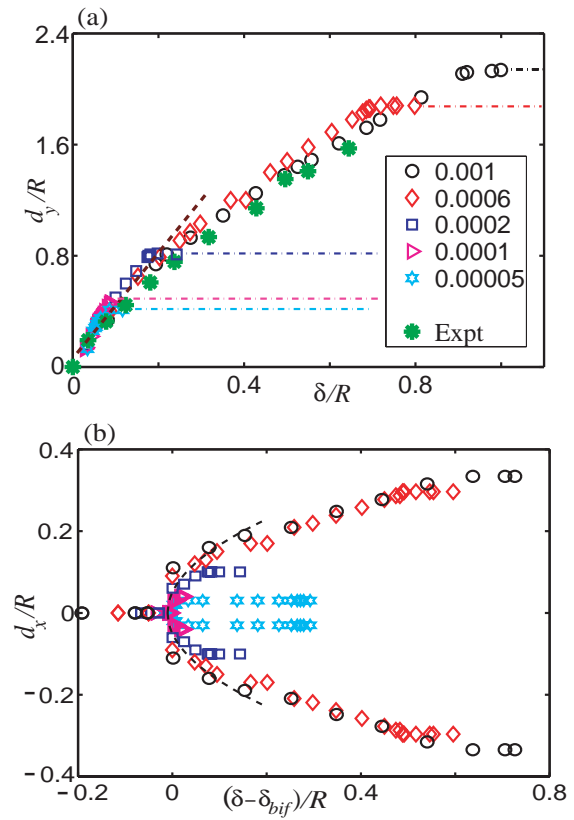}\\
\caption{\label{FEsat}(Color Online) Curvature condensation and twinning. (a) The axial location $d_{y}$ (along the symmetry axis) of the condensate as a function of normalized indentation $\delta/R$ for  different  thicknesses ($t/R$ as shown in the legend), with the dash-dotted lines showing saturation. The asterix represent experimental data from \cite{Calladine:94}. The dashed line shows linear scaling in the initial stage. (b) The defect  location $d_{x}$  as a function of the indentation follows the simple law  $d_x \sim (\delta-\delta_{bif})^{1/2}$ (dashed line) in the neighborhood of the bifurcation from one to two defects. }
\end{figure}

As the cylindrical structure develops a hinge at location $d_y$  in response to the indentation $\delta$, geometry implies that $d_y \sim \delta$. Of course as $\delta$ is increased further, we expect $d_y$ to eventually saturate at a location $d_{y, sat} \sim R$. In Fig. \ref{FEsat}a, we see that numerical simulations confirm these naive expectations: for  small indentations following the onset of localization, the location of the condensate increases linearly with $\delta$ before it eventually saturates with $d_{y,sat} \sim R$. For very thin sheets, the curvature condensate bifurcates into two condensates when the indentation $\delta > \delta_{bif}$; each of the symmetric condensates moves away from the axis of symmetry, as can be seen in Fig. \ref{FEsat}b.  The critical indentation $\delta_{bif}$ where the condensate bifurcates into two is found to follow the scaling $\delta_{bif} \sim t^{0.5}$.

\begin{figure}
\includegraphics[width=8.5cm]{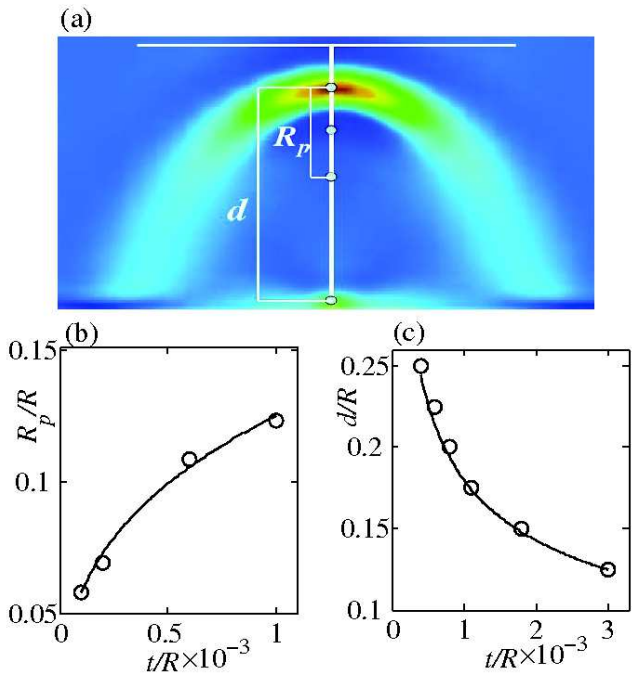}\\
\caption{\label{widthRpscaling} (Color Online) Length scales characterizing the condensate as a function of the cylinder geometry $t/R$. (a) A magnified view of the curvature condensate in Fig. \ref{curvevolve}f showing the lengthscales $R_p$ and $d$. (b) The normalized radius of curvature $R_p/R$ of the parabolic crease as a function of the normalized thickness $t/R$, at $\delta=0.08$, with the solid line  $  \sim (t/R) ^{1/3}$. (c) The normalized  location of the condensate $d/R$ as a function of  $t/R$ at $\delta/R=0.05$, with the solid line  $ \sim (t/R)^{-1/3}$. In each case, the open circles correspond to numerical simulations.}
\end{figure}

The location of the condensate is also a function of the cylinder thickness $t$; thicker, stiffer sheets lead to larger $d_{y,sat}$. In Fig.\ref{FEsat}a, we see that $d_{y,sat}$ is indeed larger for greater values of $t$ since it is energetically favorable to have the curvature condense further from the indentation.  Zooming in on the region of curvature condensation, we observe that the curvature is localized along a parabolic crescent-shaped structure (Fig.\ref{widthRpscaling}a)  similar to the core of a developable cone  \cite{Cerda:98, Boudaoud:00,Cerda:05,Liang:05}, where stress is focused anisotropically.  Our simulations show that for a given indentation, the narrowest width of the parabola $w \sim t$ occurs along the axis of symmetry, while its radius of curvature  $R_p \sim 2 t^{1/3} R^{2/3}$ (Fig.\ref{widthRpscaling}b). In Fig.\ref{widthRpscaling}c we see that the scaled location of the curvature condensate  $d/R \sim (t/R)^{-1/3}$.

We now seek a qualitative understanding of these results using the Donell-F\"{o}ppl-von Karman  equations that characterize the in and out -of-plane deformations of a weakly curved sheet \cite{Mansfield:64}. In the neighborhood of the condensate, the elastic bending energy  $E_B \approx E t^3 \left(\frac{\delta}{d w} \right)^2 {R_p}^2$, where $\delta / d w$ is the dominant component of the curvature of the shell along the cylinder axis, and $R_p^2$ is the approximate area of inversion of the cylinder due to the formation of the crease of radius $R_p$. Stretching in the vicinity of the crease is primarily due to the non-vanishing Gauss curvature of the crease $\kappa_G \approx  \frac{1}{R} \frac{\delta}{d w}$ via Gauss's Theorema Egregium. Then, we may write the stretching energy $E_S \sim Et  \gamma^2 R_p w$, where the stretching strain $\gamma \sim \sigma/Et$ in terms of the in-plane stress $\sigma \sim Et  \nabla^{-2} \kappa_G$ and $R_p w$ is the area over which this stretching arises.  This last relation follows from the compatibility relation for the stresses in the F\"{o}ppl-von Karman equation  which reads $\nabla^{2} \sigma \sim -\kappa_G$. The narrowest width of the anisotropic  crease is limited by the thickness so that $w \sim t$, consistent with our numerical simulations. Using this fact in minimizing the total energy $E_B+E_S$ yields $R_p \sim t^{1/3} R^{2/3}$  which agrees with our simulations over the range explored (Fig.\ref{widthRpscaling}b), consistent with earlier results on the size of the core of the defect in a sheet bent into a cone \cite{Cerda:98, Cerda:05,Liang:05} .  Assuming that indentation leads to an effective deformation wherein the cylinder is reflected about a parabolic boundary in the neighborhood of the localized defect, we find that  $d \approx \delta R / R_p \sim \delta (R/t)^{1/3}$,  seen in the simulations (Fig.\ref{widthRpscaling}c). For very thin shells, the curvature condensate bifurcates into two which move to a location $\pm d_x$ on either side of the axis of symmetry as the indentation increases beyond a threshold.  Indeed, symmetry considerations thus demand that this supercritical pitchfork bifurcation must follow the scaling $d_x \sim (\delta -\delta_{bif})^{1/2}$ consistent with what we observe in our numerical simulations (Fig. \ref{FEsat}b).

Our studies here are but a first step in understanding the question of defect formation and evolution in an elastic system.  By indenting the edge of a semi-cylindrical shell we followed the exploration of the geometric localization of deformation experimentally and numerically. Although we can understand some aspects of the curvature condensate, much still remains to be done in trying to quantify this phenomenon which is fairly generic in curved shells. For example, our understanding of the condensation of curvature is qualitative - we do not  know the conditions under which this would happen more generally. In addition, our description of twinning remains qualitative being based on symmetry arguments, not mechanistic ones. Finally our studies suggest possible applications to nanotechnology (e.g. deformation of nanotubes), molecular biology (e.g. virus shells), and physical chemistry (e.g. morphology of colloidal particles); indeed wherever these thin-walled structures arise.

\begin{acknowledgments} 
We thank  B. Nyugen  for help with some preliminary experiments, and M.P. Brenner, J. W. Hutchinson and  M. Weidman for useful discussions.
\end{acknowledgments}
 
\end{document}